\documentclass{PoS}

\title{New Physics in the Flavour Sector in the presence of Flavour Changing Neutral Currents}

\ShortTitle{BSMwithFCNC}
\author{\speaker{G. C. Branco and M. N. Rebelo}\thanks{
Write-up of two invited Workshop talks given separately by the two authors with
titles: ``New Physics in the
Flavour Sector in the Presence of Heavy Fermions" (GCB) and
``Minimal Flavour Violation with Two Higgs Doublets" (MNR)}\\
Centro de F\'isica Te\'orica de Part\'iculas,
Instituto Superior T\'ecnico, \\ Universidade T\'ecnica de Lisboa, Av. Rovisco Pais, 
1049-001 Lisboa, Portugal\\
        E-mail: \email{gbranco@ist.utl.pt, gustavo.branco@cern.ch, 
rebelo@ist.utl.pt, margarida.rebelo@cern.ch} }

\abstract{Flavour-Changing-Neutral-Currents (FCNC) play an important r\^ ole
in testing the Standard Model (SM) while probing the possibility of having
New Physics beyond the SM. In the SM, FCNC  are forbidden at three level, but
arise through calculable one-loop contributions. We review some of the features 
of FCNC in two examples of minimal extensions of the SM. In the first example,
we consider an extension of the SM consisting of the addition of one vector-like
quark either of the up-type ($Q=2/3$) or the down type  ($Q=- 1/3$). In this extension
there are non-vanishing but naturally suppressed Z-mediated FCNC at tree level.
In the second example, we discuss extensions of 
the SM with two Higgs doublets, without the assumption of natural flavour conservation,
giving rise to Higgs mediated FCNC. The existence of strict experimental limits on
processes sensitive to Higgs FCNC requires a strong suppression of these currents.
We present scenarios resulting from discrete symmetries where all new flavour 
structures in the quark sector are  parametrized by elements of the 
Cabibbo-Kobayashi-Maskawa  (CKM) matrix, 
together with the ratio of vacuum expectation values of the Higgs doublets in the 
Higgs basis defined by the symmetry. We extend these scenarios to the leptonic sector
with the Pontecorvo-Maki-Nakagawa-Sakata matrix playing a r\^ ole similar
to the CKM matrix in the quark sector.}

\FullConference{Proceedings of the Corfu Summer Institute 2012 "School and Workshops on Elementary Particle Physics and Gravity"\\
		September 8-27, 2012\\
		Corfu, Greece}

\begin{document}

\section{Introduction}
The Standard Model (SM) is very successful in accounting for the experimental
observations of the hadronic sector except for a few anomalies and
tensions, still to be confirmed.  In the leptonic sector we are confronted with 
a different situation.  In the SM, neutrinos are strictly massless: there are no
Dirac neutrino masses due to the absence of the righthanded
neutrino fields $\nu_{Ri}$ and no Majorana masses are generated in higher 
orders, due to exact B-L conservation. Therefore non-vanishing neutrino
masses require Physics Beyond the SM.

Extending  the SM in order to account for the observed 
leptonic mixing and neutrino masses involves novel features, not present in the 
quark sector. Even the most straightforward extension consisting of 
simply introducing righthanded neutrinos opens up the possibility
of very rich  new phenomena such as baryogenesis through leptogenesis.
By now, it has been established that in the SM it is not possible to generate the 
observed baryon asymmetry of the universe (BAU). In particular, new sources of CP
violation are required. Therefore neutrino masses and the observed BAU
provide two of the motivations to consider New Physics beyond the SM.
 
In this note,  we consider two simple extensions of the SM, where FCNC arise
at tree level, but are naturally suppressed. In the first example, described in
 Section 2, we extend the SM with vectorial isosinglet quarks which leads to Z mediated
flavour changing neutral currents (FCNC) as well as deviations from unitarity of the
Cabibbo-Kobayashi-Maskawa (CKM) matrix, in such a way that the strength of both 
effects are inter-related. Furthermore, such extensions allow for a natural 
suppression of these effects, as required by experiment.
In Section 3 we discuss a two Higgs doublet model, without natural flavour
conservation, in the Higgs sector,
where all new flavour structures in the quark sector are  parametrized 
by elements of  the CKM matrix, together with the ratio of vacuum expectation values 
of the Higgs doublets, whereas in the leptonic sector the same r\^ ole is played
by the Pontecorvo-Maki-Nagawa-Sakata (PMNS) matrix. In general two Higgs doublet 
models have Higgs mediated FCNC as well as processes mediated by a charged Higgs field
which, of course, is not present in the SM.  The novel feature of the class of models
described here is the fact that the flavour structure of FCNC only depends on
the CKM matrix and can be naturally suppressed by small CKM matrix elements.

\section{New Physics in the Flavour Sector in the Presence of Heavy Fermions}
One of the dogmas in the construction of unified gauge models is the absence of
Z-mediated tree-level flavour changing neutral currents. The origin of this dogma 
\cite{Glashow:1976nt}, \cite{Paschos:1976ay} stems from the fact that Z-mediated
FCNC, if not suppressed, lead to too large contributions to various processes
like $K^0_L \rightarrow \mu^+ \mu^-$, $K_L - K_S$ mass difference,
$K^+   \rightarrow \pi^+ \nu \overline{\nu}$, etc. One may ask the question whether
this dogma can be violated in realistic and plausible extensions of the SM. In 
this section we emphasize that this is indeed the case. This talk is based on
work done in the framework of models with vector-like quarks
\cite{Botella:2008qm}, \cite{Botella:2012ju}.
Models with vector-like quarks ( see also \cite{vectorlike} ) 
provide a framework where there are FCNC at tree level, which are naturally 
suppressed by factors of $m^2/M^2$, where $m$ and $M$ stand for the masses 
of the SM quarks and the vector-like quarks. For definitness, let us consider 
an extension of the SM where one up-type isosinglet quark $T$ is added 
to the SM spectrum \cite{Botella:2012ju}. Both $T_L$ and $T_R$ are 
isosinglets, so mass terms of the type $\overline{T_L} T_R$ , $\overline{T_L} u_{Rj}$
($j=1$ to 3 ) are $SU(2) \times U(1)$ gauge invariant and can be large. Without
loss of generality one can choose a weak basis where the down quark mass matrix is diagonal real.  In this basis, $U$ is just the $4\times 4$ unitary matrix which enters the diagonalization of the up quark mass matrix. With no loss of generality, one can also use the freedom to rephase quark fields, to choose the phases of $U$ in the following way:
\begin{equation}
\mbox{arg}(U)=\left( 
\begin{array}{cccc}
0 & {\chi ^{\prime }} & -\gamma & ...  \\ 
\pi  & 0 & 0 & ...\\ 
-\beta  & \pi +\chi  & 0 & ... \\
... & ... & ... & ... 
\end{array}
\right) 
\label{EQ:CKMphases01}
\end{equation}
where the four rephasing invariant phases are  \cite{Aleksan:1994if} ,\cite{Branco:1999fs}:
\begin{eqnarray}
 \beta\equiv \arg(-V_{cd} V^*_{cb} V^*_{td} V_{tb})&;\quad& \gamma\equiv \arg(-V_{ud} V^*_{ub} V^*_{cd} V_{cb});\nonumber\\
 \chi\equiv \arg(-V_{ts} V^*_{tb} V^*_{cs} V_{cb})&;\quad& \chi^\prime\equiv \arg(-V_{cd} V^*_{cs} V^*_{ud} V_{us}).\label{EQ:CKMphases02}
\end{eqnarray}
some authors use $\beta_s\equiv\chi$, $\phi_1\equiv\beta$ and $\phi_3\equiv\gamma$; $\chi^\prime$ is usually neglected.
It should be emphasized that independently of the dimensions of $U$, only the four rephasing invariant phases in \ref{EQ:CKMphases02} enter its $3\times 3$ sector connecting standard quarks. In the three generations SM, these four rephasing invariant phases and the nine moduli of $V_{CKM}$ are related by various exact relations \cite{Botella:2002fr} which provide a test of the SM. It can be readily verified that in the context of the SM, the phases $\chi$ and $\chi^\prime$ are small, of order $\lambda^2$ and $\lambda^4$, respectively, with $\lambda\simeq 0.2$. It has been pointed out that in the framework of models with up-type isosinglet quarks \cite{AguilarSaavedra:2004mt}, one can obtain larger values of $\chi$
The recent measurements of $\chi$ are in agreement with the SM, 
but the errors are large and it is clear that there is room for New Physics contributions,
which can be discovered once a better precision is obtained in the measurement of $\chi$.

As mentioned above, we assume that there is only one up-type isosinglet quark, which we denote T. In the mass eigenstate basis the charged and neutral current interactions can be written:
\begin{eqnarray}
\cal L_W &=&-\frac{g}{\sqrt 2}\bar {\bf u}_L \gamma^\mu V  {\bf d}_L W_\mu^\dagger+
\mbox{H.c.}~,\nonumber\\
 \cal L_Z &=&-\frac{g}{2\cos\theta_W} \left[\bar {\bf u}_L \gamma^\mu (V V^\dagger)     
{\bf u}_L-\bar {\bf d}_L \gamma^\mu 
{\bf d}_L-2\sin^2\theta_W J^\mu_{em}\right] Z_\mu~,
\label{EQ:LWLZ}
\end{eqnarray}
where 
${\bf u}=(u,c,t,T)$, ${\bf d}=(d,s,b)$,
while $V$ is a $4\times 3$ submatrix of the $4\times 4$ unitary matrix $U$ which enters the diagonalization of the up-type quark mass matrix:
\begin{equation}
 V  =  \left(\begin{array}{ccc}  
V_{ud}  & V_{us}  & V_{ub}  \\
V_{cd}  & V_{cs}  &  V_{cb}  \\
V_{td}  & V_{ts}  & V_{tb}  \\
V_{Td}  & V_{Ts}  & V_{Tb}
\end{array}\right) .
\label{EQ:CKMmatrix}
\end{equation}
It is clear from Eqs. (\ref{EQ:LWLZ}), (\ref{EQ:CKMmatrix}), that $V V^\dagger\neq 1$, which leads to FCNC in the up-quark sector.  Writing explicitly:
\begin{equation}
( V V^\dagger ) _{ij} = \delta_{ij} - U_{i4} U^*_{j4}
\end{equation}
one sees that deviations from unitarity are controlled by $U_{i4} U^*_{j4}$
The salient feature of this class of models with isosinglet quarks is that there are naturally small violations of unitarity. It is clear from Eq. \ref{EQ:CKMmatrix} that the columns of $V$ are orthogonal, while its rows are not. It can be readily verified \cite{Bento:1991ez}
that deviations of unitarity are suppressed by $m^2/M^2$, where $m$ and $M$ 
stand for the standard quark and vector-like quark masses, respectively.
At this point, it should be emphasized that there is nothing ``strange" in having small
violations of $3 \times 3$ unitarity. The leptonic mixing matrix also has small deviations 
of unitarity in the seesaw type one framework.

One may summarize some of the implications of the addition of one isosinglet
up- type vector-like quark in the following way \cite{Botella:2013bsa}:
\begin{itemize}

\item Leads to the he inclusion of a new mass eigenstate in the up sector which 
can give new contributions to amplitudes involving virtual up quarks as for example in 
kaon and B-meson mixings.

\item  Leads to a quark mixing matrix $V$ which is not $3 \times 3$ unitary
allowing for deviations of the elements $V_{ij}$  from SM values

\item Leads to moified couplings to the Z-bosons in the up-sector, including
tree level flavour changing couplings and a reduced value of the flavour 
conserving couplings.

\item  Leads to modifications in the $bd$ sector which can alleviate the existing tensions.

\end{itemize}

Next we briefly mention some of the consequences of having small deviations
of unitarity. Although our analysis is done within the framework of one isosinglet quark $T$, a good part of our results hold in a much larger class of extensions of the SM. The crucial ingredient is the presence of small violations of unitarity, independently of their origin.

In the SM, using $3 \times 3$ unitarity of $V_{CKM}$, we can derive exact relations 
between rephasing invariant  $V_{CKM}$ phases and the moduli of $V_{CKM}$. These
relations are obviously modified in the presence of an up-type vector-like quark.
As an example, let us consider the estimated value of $\chi$ in the present model.

From orthogonality of the second and third column of $V$, one obtains \cite{AguilarSaavedra:2004mt}:
\begin{equation}
 \sin\chi=\frac{ |V_{ub}| |V_{us}| } {|V_{cb}| |V_{cs}|} 
\sin(\gamma-\chi+\chi^\prime)+
\frac{|V_{Tb}| |V_{Ts}|} {|V_{cb}| |V_{cs}|} \sin(\sigma-\chi)~,
\label{EQ:32Corth}
\end{equation}
where $\sigma$ is a rephasing invariant phase, 
$\sigma\equiv \arg(V_{Ts}V_{cb}V^*_{Tb} V^*_{cs})$. 
In the SM one has, of course, $\sin\chi=\mathcal O(\lambda^2)$, since only the 
first term in Eq.~(\ref{EQ:32Corth}) is present. It is clear that in this extension of the SM one
may obtain a significant deviation from the SM value. One may obtain a significant 
enhancement if $|V_{Tb} V_{Ts}|$  is not too small or one may obtain a 
suppression of $\chi$ if the two terms in Eq.~(\ref{EQ:32Corth}) have opposite signs.

This model has FCNC in the up sector and in particular one has couplings of
the type $\bar{c_L} \gamma^\mu t_L Z_{\mu}$ which are proportional
to $|u_{24} u_{34}|$, which measures deviations of orthogonality of the
second and third rows of $V$. Provided $|u_{24} u_{34}|$ is not too small, one
may have rare top decays $t \rightarrow cZ$ at rates which can be observed 
at the LHC. In this model one also has $Z$ couplings to  $\bar{c_L} \gamma^\mu u_L$
at tree level \cite{Branco:1995us} . In order for these couplings to be able
to account for the observed size of $D^0 - \bar{D^0}$ mixing, the size of 
$|u_{14} u_{24}|$ has to be of order $\lambda^5$ \cite{Golowich:2007ka} .

It has also been pointed out that \cite{Botella:2012ju} that in the framework 
of this model one has the potential for solving the tension between experimental 
values of $A_{J/\Psi K_S}$ and Br($B^+ \rightarrow \tau^+ \nu_\tau$)
with respect to SM expectations. One may also have important deviations
from the SM in observables in the bd sector like the semi-leptonic
asymmetry $A^d_{SL}$, $B^0_d \rightarrow \mu^+ \mu^-$ and
$A^s_{SL} - A^d_{SL}$. Other potential places where NP can show up include
$ A_{J/\Psi \phi}$, $\gamma$, $K^0_L \rightarrow \pi^0 \nu \bar \nu$,
$D^0 \rightarrow \mu^+ \mu^-$ \cite{Botella:2013bsa}.

\section{Minimal Flavour Violation with Two Higgs Doublets}
The flavour structure of Yukawa couplings is not constrained by gauge
invariance. In the SM all flavour changing transitions are mediated by
charged weak currents with flavour mixing controlled by   $V_{CKM}$, the
Cabibbo-Kobayashi-Maskawa matrix. Models with two 
Higgs doublets \cite{Branco:2011iw},  \cite{Gunion:1989we} 
have potentially large Higgs FCNC. The existence of
strict limits on FCNC processes requires a mechanism of suppression.
The elimination of tree level FCNC is accomplished, for instance, in 
the context of natural flavour conservation \cite{Glashow:1976nt}
through a discrete symmetry such that only one Higgs doublet couples and  
gives mass to each fermionic  sector.  An alternative proposal is the Aligned two 
Higgs doublet model \cite{Pich:2009sp}.
An alternative idea, put forward in the early nineties,  
is to have tree level Higgs mediated FCNC suppressed by
small factors given in terms of small entries of the $V_{CKM}$ 
matrix \cite{Ant+hall}, \cite{Joshipura:1990pi}. The first 
models of this type with no ad-hoc assumptions. obtained from a symmetry, 
were proposed by Branco, Grimus and Lavoura \cite{Branco:1996bq} (BGL).  
Later on, we have generalized BGL models \cite{Botella:2009pq}, and extended the 
idea to the leptonic  sector \cite{Botella:2011ne} as reported 
in this talk. In the early year two thousands the designation  Minimal Flavour Violation (MFV)
was coined \cite{D'Ambrosio:2002ex}, \cite{Cirigliano:2005ck},
referring to extensions of the SM model where the breaking of the 
large $U(3)^5$ flavour symmetry of the gauge sector  is completely determined by Yukawa couplings, as it is the case in the SM.  The definition  requires, in addition,  that
the top quark Yukawa couplings should play a special r\^ ole. Due to this requirement,
not all BGL implementations, which are presented below,  fall into the category 
of models  considered as being of MFV type, only a specific 
example out of the six possible BGL models is recognized as such by authors of 
the definition \cite{Buras:2010mh}.  An interesting alternative definition of MFV 
in the context of two Higgs doublet models was given  and discussed 
in a recent work \cite{Dery:2013aba}. A feature common to all  these models 
is the fact that the flavour structure of the quark sector is expressed  in terms 
of entries of the $V_{CKM}$ matrix.  A distinctive feature of  BGL models  
is that they are obtained from a  global Abelian symmetry.

In order to fix our notation,
we specify the Yukawa interactions, starting with the quark sector:
\begin{equation}
L_Y = - \overline{Q^0_L} \ \Gamma_1 \Phi_1 d^0_R - \overline{Q^0_L}\  
\Gamma_2 \Phi_2 d^0_R - \overline{Q^0_L} \ \Delta_1 \tilde{ \Phi_1} u^0_R - 
\overline{Q^0_L} \ \Delta_2 \tilde{\Phi_2} u^0_R \ + \mbox{h. c.}
\label{1e2}
\end{equation} 
where $\Gamma_i$ and $\Delta_i$ denote the Yukawa couplings of the 
lefthanded quark doublets $Q^0_L$ to the righthanded quarks $d^0_R$,
$u^0_R$ and the Higgs doublets $\Phi_j$. The quark mass matrices generated
after spontaneous gauge symmetry breaking are given by:
\begin{eqnarray}
M_d = \frac{1}{\sqrt{2}} ( v_1  \Gamma_1 +
                           v_2  e^{i \alpha}   \Gamma_2 ), \quad 
M_u = \frac{1}{\sqrt{2}} ( v_1  \Delta_1 +
                           v_2  e^{-i \alpha} \Delta_2 ),
\label{mmmm}
\end{eqnarray}
where $v_i \equiv |<0|\phi^0_i|0>|$ and $\alpha$ denotes the relative phase 
of the vacuum expectation values (vevs) of the neutral components of 
$\Phi_i$. 
The matrices $M_d$,  $M_u$ are diagonalized by the usual bi-unitary transformations:
\begin{eqnarray}
U^\dagger_{dL} M_d U_{dR} = D_d \equiv {\mbox diag}\ (m_d, m_s, m_b) 
\label{umu}\\
U^\dagger_{uL} M_u U_{uR} = D_u \equiv {\mbox diag}\ (m_u, m_c, m_t)
\label{uct}
\end{eqnarray}
The neutral and the charged Higgs interactions obtained from the quark 
sector of  Eq.~(\ref{1e2}) are of the form
\begin{eqnarray}
{\mathcal L}_Y (\mbox{quark, Higgs})& = & - \overline{d_L^0} \frac{1}{v}\,
[M_d H^0 + N_d^0 R + i N_d^0 I]\, d_R^0  - \nonumber \\
&-& \overline{{u}_{L}^{0}} \frac{1}{v}\, [M_u H^0 + N_u^0 R + i N_u^0 I] \,
u_R^{0}  - \label{rep}\\
& - &  \frac{\sqrt{2} H^+}{v} (\overline{{u}_{L}^{0}} N_d^0  \,  d_R^0 
- \overline{{u}_{R}^{0}} {N_u^0}^\dagger \,    d_L^0 ) + \mbox{h.c.} \nonumber 
\end{eqnarray}
where $v \equiv \sqrt{v_1^2 + v_2^2} \approx \mbox{246 GeV}$,  
and $H^0$, $R$ are orthogonal combinations of the fields  $\rho_j$,
arising when one expands  \cite{Lee:1973iz}  the neutral scalar fields
around their vacuum expectation values, $ \phi^0_j =  \frac{e^{i \alpha_j}}{\sqrt{2}} 
(v_j + \rho_j + i \eta_j)$, choosing $H^0$ in such a way
that it has couplings to the quarks which are proportional
to the mass matrices, as can be seen from Eq.~(\ref{rep}). 
Similarly, $I$ denotes the linear combination
of $\eta_{j}$ orthogonal to the neutral Goldstone boson. 
The matrices $N_d^0$,  $N_u^0$  
are given by:
\begin{eqnarray}
N_d^0 = \frac{1}{\sqrt{2}} ( v_2  \Gamma_1 -
                           v_1 e^{i \alpha} \Gamma_2 ), \quad 
N_u^0 = \frac{1}{\sqrt{2}} ( v_2  \Delta_1 -
                           v_1 e^{-i \alpha} \Delta_2 )
\end{eqnarray}
The flavour structure of the quark sector of  two Higgs doublet models
is thus fully specified in terms of
the four matrices $M_d$,  $M_u$, $N_d^0$,  $N_u^0$.  
In terms of the quark mass eigenstates $u, d$, the Yukawa couplings are:
\begin{eqnarray}
L_Y & = & - \frac{\sqrt{2} H^+}{v} \bar{u} \left(
V N_d \gamma_R - N^\dagger_u \ V \gamma_L \right) d +  \mbox{h.c.} - 
\frac{H^0}{v} \left(  \bar{u} D_u u + \bar{d} D_d \ d \right) - 
\nonumber \\
& - & \frac{R}{v} \left[\bar{u}(N_u \gamma_R + N^\dagger_u \gamma_L)u+
\bar{d}(N_d \gamma_R + N^\dagger_d \gamma_L)\ d \right] + \\
& + & i  \frac{I}{v}  \left[\bar{u}(N_u \gamma_R - N^\dagger_u \gamma_L)u-
\bar{d}(N_d \gamma_R - N^\dagger_d \gamma_L)\ d \right]
\nonumber
\end{eqnarray}
with $\gamma_L = (1 - \gamma_5)/2$,
$\gamma_R = (1 + \gamma_5)/2$ and where $V$ stands for the $V_{CKM}$ matrix.
The matrices $N_d$ and $N_u$ are:
\begin{equation}
N_d =  U^\dagger_{dL} N_d^0 U_{dR},  \qquad N_u =  U^\dagger_{uL} N_u^0 U_{uR}
\end{equation}
Comparison with Eqs. (\ref{umu}), (\ref{uct}) shows that the matrices $N_d^0$,  
$N_u^0$ transform in the same way as the matrices  $M_d$,  $M_u$ under unitary transformations of the quark fields. The physical neutral Higgs fields are combinations 
of  $H^0$, $R$ and $I$. Flavour changing neutral currents are controlled by $N_d$ and 
$N_u$. For generic two Higgs doublet models $N_d$, $N_u$ are non-diagonal 
arbitrary.

In order to obtain a structure for the matrices $\Gamma_i$ and $\Delta_i$ such that 
the the strength of the tree level FCNC is completely controlled by 
$V_{CKM}$, Branco, Grimus and Lavoura (BGL)  imposed the
following symmetry on the quark and scalar sector of the Lagrangian 
\cite{Branco:1996bq}:
\begin{equation}
Q_{Lj}^{0}\rightarrow \exp {(i\tau )}\ Q_{Lj}^{0}\ ,\qquad
u_{Rj}^{0}\rightarrow \exp {(i2\tau )}u_{Rj}^{0}\ ,\qquad \Phi
_{2}\rightarrow \exp {(i\tau )}\Phi _{2}\ ,  \label{S symetry up quarks}
\end{equation}
where $\tau \neq 0, \pi$, with all other quark fields transforming 
trivially under the symmetry. The index $j$ can be fixed as either
1, 2 or 3. Alternatively the symmetry may be chosen as:
\begin{equation}
Q_{Lj}^{0}\rightarrow \exp {(i\tau )}\ Q_{Lj}^{0}\ ,\qquad
d_{Rj}^{0}\rightarrow \exp {(i2\tau )}d_{Rj}^{0}\ ,\quad \Phi
_{2}\rightarrow \exp {(- i \tau)}\Phi _{2}\ .  \label{S symetry down quarks}
\end{equation} \\
The symmetry given by Eq.~(\ref{S symetry up quarks}) leads to Higgs FCNC
in the down sector, whereas the symmetry specified by 
Eq.~(\ref{S symetry down  quarks}) leads to Higgs FCNC in the up sector. 
In the case of the symmetry given by
Eq.~(\ref{S symetry up quarks}), for $j=3$ 
there are FCNC in the down sector controlled by the matrix  
$N_d$  given by   \cite{Branco:1996bq}
\begin{equation}
(N_d)_{ij} \equiv  \frac{v_2}{v_1} (D_d)_{ij} - 
\left( \frac{v_2}{v_1} +  \frac{v_1}{v_2}\right) 
(V^\dagger_{CKM})_{i3} (V_{CKM})_{3j} (D_d)_{jj}\ . \label{24}
\end{equation}
whereas, there are no FCNC in the up sector and the coupling matrix
of the up quarks to the $R$ and $I$ fields is of the form:
\begin{equation}
N_u = - \frac{v_1}{v_2} \mbox{diag} \ (0, 0, m_t) +  \frac{v_2}{v_1}
\mbox{diag} \ (m_u, m_c, 0)\ . \label{25}
\end{equation}
It is clear that BGL models are very constrained. Only one new parameter,
not present in the SM, 
appears in the flavour sector, that is the ratio $\tan \beta = v_2/v_1$. 
As a result of the imposed symmetry  the Higgs potential, together
with a soft symmetry breaking term, required in order to avoid an ungauged
accidental continuos symmetry,  has seven  parameters which can be chosen
to be real, without loss of generality.  The Higgs sector does not violate CP
neither explicitly nor spontaneously. The seven independent  parameters of
the potential determine the masses of the four Higgs fields, $\tan \beta$,
the quantity $v \equiv \sqrt{v_1^2 + v_2^2} $ and the mixing among
$H^0$ and $R$, which is supposed to be small due to the fact that the
Higgs field discovered at the LHC \cite{Aad:2012tfa}, \cite{Chatrchyan:2012ufa},
behaves very much like a SM Higgs field. The study of the phenomenological implications
of this class of models is underway. This requires the specification of the leptonic
sector. For Dirac neutrinos the extension is straightforward in analogy to
the quark sector. The case of Majorana type neutrinos is more involved.

In terms of the low energy effective theory for Majorana neutrino masses, a
priori, it looks more difficult to implement MFV. However, this can be done
by imposing a $Z_4$ symmetry to the  effective Lagrangian as presented in 
Ref.~\cite{Botella:2011ne} . In the seesaw case, with the introduction of
three righthanded neutrinos the leptonic part of Yukawa couplings
and invariant mass terms can then be written: 
\begin{eqnarray}
{\mathcal L}_{Y +\mbox{mass}} & = & - \overline{L^0_L} \ \Pi_1 \Phi_1 l^0_R - 
\overline{L^0_L}\  
\Pi_2 \Phi_2 l^0_R - \overline{L^0_L} \ \Sigma_1 \tilde{ \Phi_1} \nu^0_R - 
\overline{L^0_L} \ \Sigma_2 \tilde{\Phi_2} \nu^0_R \  + \nonumber \\
& + & \frac{1}{2}  {\nu_R^{0}}^T C^{-1} M_R \nu_R^{0} +
\mbox{h.c.}\ .
\label{1e3}
\end{eqnarray}
The matrix $M_R$ stands for the
righthanded neutrino Majorana mass matrix. The leptonic mass matrices
generated after spontaneous gauge symmetry breaking are given by: 
\begin{eqnarray}
m_l = \frac{1}{\sqrt{2}} ( v_1 \Pi_1 + v_2 e^{i \theta} \Pi_2 )\ , \quad m_D = 
\frac{1}{\sqrt{2}} ( v_1 \Sigma_1 + v_2 e^{-i \theta} \Sigma_2)\ .
\label{bbbb}
\end{eqnarray}
The neutral Higgs interactions with the fermions, obtained from
Eq.~(\ref{1e3}) can be written:
\begin{eqnarray}
{\mathcal L}_Y (\mbox{neutral, lepton})& = & - \overline{l_L^0} \frac{1}{v} \,
[m_l H^0 + N_l^0 R + i N_l^0 I]\, l_R^0  + \nonumber \\
&-& \overline{{\nu}_{L}^{0}} \frac{1}{v}\, [m_D H^0 + N_\nu^0 R + i N_\nu^0 I]\, 
\nu_R^{0} + \mbox{h.c.}\ ,
\label{neu}
\end{eqnarray}
with
\begin{eqnarray}
N_l^0 = \frac{v_2 }{\sqrt{2}} \Pi_1  - \frac{v_1 }{\sqrt{2}} 
e^{i \theta} \Pi_2\ ,  \\
N_\nu^0 = \frac{v_2 }{\sqrt{2}} \Sigma_1  - \frac{v_1 }
{\sqrt{2}} e^{-i \theta} 
\Sigma_2 \ .
\end{eqnarray}
There is a new feature in the seesaw framework 
due to the fact that in the neutrino sector the light 
neutrino masses are not obtained from the diagonalization of $m_D$. 
In general the couplings of  Eq.~(\ref{neu}) lead to arbitrary scalar 
FCNC at tree level. In order for these couplings to be completely controlled 
by the PMNS matrix we introduce the following 
$Z_4$ symmetry on the Lagrangian:
\begin{equation}
L^0_{L3} \rightarrow \exp{(i \alpha)}\  L^0_{L3}\ , \qquad
\nu^0_{R3} \rightarrow \exp{(i 2\alpha)} \nu^0_{R3}\ , \qquad
\Phi_2   \rightarrow \exp{(i \alpha)} \Phi_2\ , \label{bgl}
\end{equation}
with $\alpha = \pi/2 $ and all other fields transforming
trivially under $Z_4$. The most general matrices $\Pi_i$, $\Sigma_i$ and $M_R$ 
consistent with this $Z_4$ symmetry have the following structure:
\begin{eqnarray}
 \Pi_1 & = & \left[\begin{array}{ccc}  
\times  & \times & \times \\
\times & \times &  \times \\
0 & 0 & 0 
\end{array}\right]\ , \qquad
 \Pi_2   =  \left[\begin{array}{ccc}  
0 & 0 & 0  \\
0 & 0 & 0 \\
\times & \times &  \times 
\end{array}\right]\ , \label{gam}\\
 \Sigma_1  & = & \left[\begin{array}{ccc}  
\times  & \times & 0 \\
\times & \times &  0 \\
0 & 0 & 0 
\end{array}\right]\ , \qquad 
 \Sigma_2   =  \left[\begin{array}{ccc}  
0  & 0 & 0 \\
0 & 0 &  0 \\
0 & 0 & \times
\end{array}\right]\ , \qquad 
M_R   =   \left[\begin{array}{ccc}  
\times  & \times & 0 \\
\times & \times &  0 \\
0 & 0 & \times 
\end{array}\right]\ ,
\label{del}
\end{eqnarray}
where $\times$ denotes an arbitrary entry while the zeros are imposed
by the symmetry $Z_4$. Note that the choice of $Z_4$ is crucial 
in order to guarantee $M_{33}\neq 0$ and thus a non-vanishing
$\det M_R$. In this case there are flavour changing neutral currents in the
charged leptonic sector given by:
\begin{equation}
(N_l)_{ij} \equiv  ({U_l}_L^\dagger \ N_l^0 \ {U_l}_R)_{ij} 
= \frac{v_2}{v_1} (D_l)_{ij} - 
\left( \frac{v_2}{v_1} +  \frac{v_1}{v_2}\right) 
(U^\dagger_{\nu})_{i3} (U_{\nu})_{3j} (D_l)_{jj}\ . \label{64}
\end{equation}
$U_{\nu}$ is the PMNS matrix.
In the neutrino sector we have three light and three heavy neutrinos. The
light-light Higgs mediated neutral currents are flavour diagonal. On the other hand Higgs
mediated light-heavy and heavy-heavy neutrino couplings can be parametrized 
\cite{Botella:2011ne} 
in terms of neutrino masses and the orthogonal complex matrix of the
Casas and Ibarra parametrization \cite{Casas:2001sr}. This matrix plays an important 
r\^ ole for leptogenesis \cite{Fukugita:1986hr}. In the context of seesaw 
the masses of heavy neutrinos are many orders of magnitude above the TeV
scale, therefore processes involving heavy neutrinos are not relevant for 
low energy physics.

\section*{Acknowledgements}
The authors wish to thank the local Organizing Committee and in special 
George Zoupanos for the very stimulating Workshop and for the warm 
hospitality in Corfu.
The works reported in this workshop had been  partially supported by Funda\c c\~ ao 
para a Ci\^ encia e a Tecnologia (FCT, Portugal) through the projects
CERN/FP/83503/2008 and CFTP-FCT Unit 777 (PEst-OE/FIS/UI0777/2011)
which are partially funded through POCTI (FEDER),  by Marie Curie RTN
MRTN-CT-2006-035505,  and ITN "UNILHC" PITN-GA-2009-237920,
by  Accion Complementaria Luso-Espanhola
PORT2008--03, by European FEDER and FPA-2008-04002-E/PORTU by
Spanish  MICINN under grant FPA2008--02878 and GVPROMETEO 2010-056.
More recently we would like to aknowledge the support of CERN/FP/123580/2011.

\end{document}